\documentclass[prl,twocolumn,showpacs]{revtex4}
\usepackage{graphicx}
\usepackage{epsfig}
\usepackage{amssymb,latexsym,amsmath}

\usepackage{ulem}
\usepackage{color}

\begin{document}

\title{ Time-dependent thermoelectric transport for nanoscale thermal machines} 

\author{A.-M. Dar\'{e}}\email{Anne-Marie.Dare@univ-amu.fr} 
\author{P. Lombardo}
\affiliation{ Aix-Marseille Universit\'e, CNRS, IM2NP UMR 7334, \small \it  13397, Marseille, France}

\date{\today}

\begin{abstract}
We analyze an electronic nanoscale thermal machine driven by time-dependent environment: besides bias and gate voltage variations, 
we consider also the less prevailing time modulation of the couplings between leads and dot. 
We provide energy and heat current expressions in such situations, as well as expressions for the power exchanged between the dot+leads system and its outside. Calculations are made in the Keldysh
nonequilibrium Green's function framework. 
We apply these results to design a cyclic refrigerator, 
circumventing the ambiguity of defining energy flows between subsystems in the case of strong coupling.
For fast lead-dot coupling modulation, we observe transient currents which cannot be ascribed to charge tunneling.

\pacs{73.63.--b,73.50.Lw,05.70.Ln }
\end{abstract}

\maketitle

\section{Introduction}

Since the early days of thermodynamics, thermal machines have been fascinating objects. Their very applied aspect was not an obstacle to infer
quite abstract ideas~\cite{Carnot}.
Since then, substantial progress has been made, thanks primarily to statistical physics,
particularly concerning the heat concept. 
However, this is not the whole story. 
Progress is still called for in the context of mesoscopic and nanoscale systems in which, quantum effects, the hybridization of subsystems, and important fluctuations are inevitable. 
The expected advances are maybe within reach nowadays due to recent significant breakthroughs in experiments~\cite{Pekola15,Brantut13} that open opportunities for testing and developing 
new theoretical approaches. 

Besides, in the context of energy harvesting, thermoelectricity has come back to the forefront of research. In a famous paper~\cite{Hicks}, it was shown that device 
nanostructuration can be an advantage for performance.
However, confinement in nanosize electronic devices requires a proper handling of quantum properties.

Nanoscopic thermoelectric devices have been and are still the subject of numerous studies.
Even the simplest ones consisting of an effective resonant-level dot coupled with two reservoirs, have been investigated from many different perspectives. 
For example, it has been shown that subtle effects such as Kondo physics emerging from Coulomb repulsion, may have a beneficial impact on thermoelectric performance~\cite{Costi10,Azema12,Donsa14}. 
Besides, simple and fundamental questions are still open or have been recently addressed in these devices: one can cite the issue addressing efficiency at maximum power \cite{Nakpathomkun10}, one can also mention the work by Whitney~\cite{Whitney14,Whitney15}, which proves the existence of a quantum bound on the output power of thermal machines.
In parallel to their static properties, interesting dynamical issues concerning nanoscale thermoelectric devices have been raised recently. Time varying bias and gate voltages~\cite{Arrachea07,Rey07,EspositoEPL2010,Liu12,crepieux,Ludovico14,Ludovico14bis,Zhou15}, or even temperature bias~\cite{Chirla14,Brandner15} have been considered.

To probe the dynamics of the device, the time-dependent dot-lead hybridization $\Gamma(t)$ has also been widely operated, essentially for charge pumping, i.e.
outside the thermoelectric 
context (see, for example, Refs.~\cite{JauhoWingreenMeir1994,schmidt08,Croy12,Vovchenko14,Eissing15,Kaestner15}). 
However, in the latter field, modulation of this parameter has been relatively seldom considered, except in few pioneering 
works~\cite{EspositoPRE,Juergens13,Torfason13,Ludovico15}. In these studies, (except the last one which is a linear response study of systems under adiabatic ac driving),
currents were evaluated using the master equation approach. Yet, this technique, compared to the one employed in the present paper, suffers from some drawbacks, as for instance, slow driving and dot-lead weak coupling restrictions~\cite{Haupt13}. 
To overcome these limitations, we evaluate the energy and heat currents with a nonequilibrium Green's function approach (NEGF).
Even if the NEGF approach can incorporate the Coulomb interaction (not considered in this paper) through a self-energy, in that case, approximations are needed to evaluate the currents.
\begin{figure}[h!]
\begin{center}
\includegraphics[width=.5\textwidth]{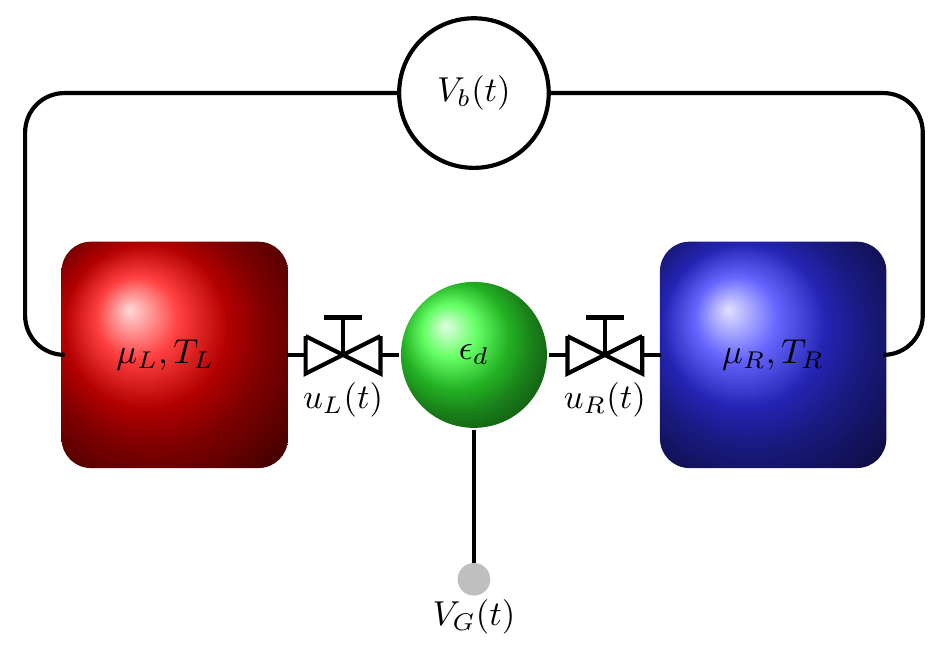}
\caption{(Color online) Schematic representation of the device: the bias voltage $V_b(t)$, the gate voltage $V_G(t)$, as well as hybridizations between dot and leads can be modulated in time.}
\label{Dispo}
\end{center}
\end{figure}

In this paper, we explore a simple device (Fig. 1) consisting of a one-level non correlated spin-degenerate dot coupled with two leads. The leads are electronic reservoirs characterized by different temperatures and chemical potentials. 
This simple thermal machine, which could be operated in the static regime, is driven by time-dependent parameters: gate and bias voltages, 
as well as hybridizations between the dot and the leads can vary in an arbitrary manner in time. 
Using the Keldysh NEGF approach, which can be solved exactly in the absence of Coulomb interaction, we evaluate the electric and thermal currents without using the slow driving hypothesis or the weak dot-lead coupling approximation. For strong hybridization between the dot and the reservoirs, there is some ambiguity in defining the energy current flowing from a lead~\cite{Ludovico14,EspositoPRL15}; indeed, the lead hybridizes with the dot and even with the second lead, thus, in some sense, looses its individuality. This has no consequence in the stationary regime, because of no energy or charge accumulations in the contact regions. But it is not so simple in the time-dependent regime. 
The situation is even trickier for the definition of heat and work exchanged between subsystems~\cite{EspositoPRL15}, for which the difficulty persists even in the static regime.
To circumvent these difficulties, we consider a cycle linking disconnected subsystem states, then, when integrated over a cycle, some vagueness disappears. 
We apply the present general expressions to a refrigerator carefully including in the balance and in the thermal machine performance, all electric works supplied to the machine.

The paper is organized as follows: after an introduction to the model, and definitions of  charge, energy, and heat currents, the energy balance along a cycle is detailed. 
Then, expressions for the charge and energy currents are given  in the Keldysh NEGF approach. After a short section about the efficiency and coefficient of performance (COP),
we  illustrate our results in the case of a time-driven refrigerator, and then conclude.
Three appendices give details about the analytical technique, and the relation between NEGF results and perturbation theory.

\section{The model} 

\subsection{Hamiltonian}
When connected together, the Hamiltonian describing the left and right leads ($\alpha =$ L,R) and the one-orbital dot, reads in standard notation
\begin{equation}
\hat{H} = \hat{H}_{D}+ \sum_{\alpha} \hat{H}_{\alpha} +\sum_{\alpha} \hat{H}_{T \alpha} \ ,
\label{Hamitonien}
\end{equation}
where 
\begin{eqnarray}
& &\hat{H}_{D} =\sum_{ \sigma} \epsilon_{d } (t) d^{\dagger}_{\sigma} d_{ \sigma}  \ ,\nonumber \\
& &\hat{H}_{\alpha}  =  \sum_{k, \sigma} \varepsilon_{k \alpha } (t) c^{\dagger}_{k \alpha \sigma} c_{k \alpha \sigma}  \ , \nonumber\\
& &\hat{H}_{T \alpha} = \sum_{k, \sigma} (V_{k \alpha } (t) c^{\dagger}_{k \alpha \sigma} d_{ \sigma} +{\mathrm h.c.})  \ .
\label{Hdetail}
\end{eqnarray}
In a quantum nanoscopic thermal machine as considered in this paper, there is no clear boundary between the central system and the thermal baths: indeed, even in the absence of electronic interactions, quantum correlations mix the central system and the reservoirs through the hybridization parameters $V_{k \alpha }$. In the so-called weak coupling situation, a simpler representation is restored due to the smallness of these quantities.

In the present paper, the dot level $\epsilon_{d }$, as well as the dispersion in the leads, $\varepsilon_{k \alpha }$, and also the hybridization between the dot and the leads, $V_{k \alpha }$,  will be allowed to depend on time as follows~\cite{JauhoWingreenMeir1994} 
\begin{eqnarray}
\epsilon_{d} (t) & =& \epsilon^0 +\Delta (t) \ , \nonumber \\
\varepsilon_{k \alpha} (t) & = & \varepsilon^0_{k \alpha} +\Delta_{\alpha }(t)\ , \nonumber \\
V_{k \alpha } (t) &= &u_{\alpha }(t) \  V_{\alpha} (\varepsilon^0_{k \alpha }) \ .
\end{eqnarray}
Without loss of generality, $ u_{\alpha }(t)$ will be supposed real~\cite{Splettstoesser05}.
$\epsilon^0 + \Delta(t) $ is related to the gate voltage, and the difference $[\mu_R -\Delta_R(t)]-[ \mu_L - \Delta_L(t)] $ is the bias voltage, with 
$\mu_{\alpha}$ the chemical potential of the lead $\alpha$.  
In the context of thermal properties, modulating the hybridization parameters has scarcely been explored. However, it can be realized experimentally at quite high frequencies~\cite{Kaestner08,Blumenthal07}.

\subsection{Currents}

In the Heisenberg representation,  we have for the time derivative of the Hamiltonian
\begin{eqnarray}
{\hat{\dot{H}}} &=& \sum_{\sigma} \dot{\epsilon}_d(t)  d^{\dagger}_{ \sigma}(t) d_{ \sigma}(t) +\sum_{\alpha,k, \sigma}\dot{\varepsilon}_{k \alpha } (t)  c^{\dagger}_{k \alpha \sigma}(t) c_{k \alpha \sigma}(t)  \nonumber \\
& + & \sum_{\alpha, k, \sigma} (\dot{V}_{k \alpha} (t) c^{\dagger}_{k \alpha \sigma}(t) d_{ \sigma}(t) +\mathrm {h.c.}) \ .
\label{Hpoint}
\end{eqnarray}
Indeed, due to $[\hat{H},\hat{H}]=0$, even in Heisenberg representation, only the time derivative of the parameters appears in $\hat{\dot{H}}$. 
In order to alleviate the preceding formula, the representation is not explicit. Henceforth, all operators will be in the Heisenberg picture.
The three right terms in Eq. (\ref{Hpoint}), after time integration, are identified as work~\cite{Gemmer}, exchanged between the whole system (dot+leads) and its outside. The first one is exchanged via the gate which is capacitively coupled to the dot. The second one is exchanged with the apparatus creating the time-dependent part of the bias voltage applied to the leads. The last one results from the time modulation of the hybridization parameters. Note that this is not the whole story concerning the work: due to its static nature, the work corresponding to the static bias voltage does not appear in Eq. (\ref{Hpoint}).

Choosing the convention of positive electric and energy currents for charge or energy entering the dot, we define the following current operators  $(e>0)$:
\begin{eqnarray}
\hat{J}_{\alpha}^e(t) & = &  e \hat{J}_{\alpha}^N(t) = -e \frac{d \hat{N}_{\alpha}}{dt}  \ , \\
\hat{J}_{\alpha}^E(t)  & = & - \frac{d \hat{H}_{\alpha}}{dt} \ ,
\label{currents}
\end{eqnarray}
with $ \hat{N}_{\alpha} = \sum_{k, \sigma}  c^{\dagger}_{k \alpha \sigma} c_{k \alpha \sigma} $. 
The designation of $\hat{J}_{\alpha}^E(t)$ as instantaneous energy current extracted from the $\alpha$ lead may be questionable (see Ref.~\cite{Ludovico14}, for example), one may also associate to $\hat{H}_{\alpha}$ a part of $\hat{H}_{T \alpha}$. 
However, the ambiguity will be lifted in the forthcoming application when considering a cyclic protocol between disconnected subsystem states. 

Due to the explicit time dependence of $\hat{H}_{\alpha}$, the energy current $\hat{J}_{\alpha}^E(t)$ includes a work contribution in addition to a thermal one and a chemical one. From the first law of thermodynamics we have for the heat current  extracted from the $\alpha$ lead
\begin{equation}
\hat{J}^Q_{\alpha}(t) =\hat{J}^E_{\alpha}(t)+\sum_{k, \sigma} \dot{\Delta}_{ \alpha } (t) c^{\dagger}_{k \alpha \sigma}(t) c_{k \alpha \sigma}(t)  -\mu_{\alpha} \hat{J}^N_{\alpha}(t) \ .
\label{heatcurrent}
\end{equation}

\subsection{Balances}

Due to charge conservation, the instantaneous balance for the electric current reads 
\begin{equation}
\sum_{\alpha} \hat{J}_{\alpha}^e(t) = e \sum_{\sigma} \hat{\dot{n}}_{ \sigma} (t) \ ,
\end{equation}
with ${\hat{n}}_{ \sigma} =  d^{\dagger}_{ \sigma} d_{ \sigma}$, 
using the notation $\hat{\dot{A}} = \frac{d \hat{A}}{d t} =  \frac{\partial \hat{A}} {\partial t}  + \frac{i}{\hbar} [ \hat{H}, \hat{A}]$.
In contrast, the energy is not conserved and the power balance can be expressed as 
\begin{equation}
\hat{\dot{H}}_{D}(t)+\sum_{\alpha} \hat{\dot{H}}_{T \alpha}(t) = \sum_{\alpha} \hat{J}_{\alpha}^Q(t)+
\hat{\mathcal{P}}_G(t)+\ \hat{\mathcal{P}}_{\Gamma}(t)+\hat{\mathcal{P}}_{ch}(t) \ ,
\label{ebilan}
\end{equation}
with 
\begin{eqnarray}
\hat{\mathcal{P}}_G(t) & = & \dot{\Delta}(t)\sum_{\sigma} \hat{n}_{\sigma}(t) \ ,\nonumber \\
\hat{\mathcal{P}}_{\Gamma} (t)& = & \sum_{\alpha}\hat{\mathcal{P}}_{\alpha \Gamma}(t) =\sum_{\alpha, k, \sigma} (\dot{V}_{k \alpha } (t) c^{\dagger}_{k \alpha \sigma}(t) d_{ \sigma}(t) +\mathrm{h.c.}) \ ,\nonumber \\
\hat{\mathcal{P}}_{ch}(t) & =&  \frac 1 e \sum_{\alpha}  \mu_{\alpha} \hat{J}^e_{\alpha} \ . 
\label{powers}
\end{eqnarray}
These three terms correspond to power exchanged with the outside respectively through time modulation of the gate voltage $(\hat{\mathcal{P}}_G)$, through time modulation of hybridization parameters  $(\hat{\mathcal{P}}_{\Gamma})$, and through the time-independent bias voltage  $(\hat{\mathcal{P}}_{ch})$, the last one is sometimes called chemical power. In the stationary case, it reduces to
$ \frac{\mu_{L} - \mu_{R}}{e} \hat{J}^e_{L}$, while the other two disappear. 
It may be tempting to read Eq.(\ref{ebilan}) as a balance concerning some extended dot (dot + contact regions). However, this would be questionable because powers are exchanged between the whole system (including the dot and the two leads) and its outside. 

Due to the ambiguity in defining the energy flowing from the leads,
the interpretation of $<\hat{J}^Q_{\alpha}(t)>$  in terms of heat currents extracted from the respective reservoirs  is also questionable. 
However, when integrated along a cycle linking disconnected subsystem states, the energies extracted from the leads are no more ambiguous. This is not true for the heat, which is not a state function except in the special case when the heat and energy currents coincide (see later our setup).

In a forced regime, for a periodic modulation of time-dependent parameters,  after the fading of the initial state influence~\cite{initial}, the currents and dot occupancy become periodic in time.
When integrated along a cycle of duration $\tau$, in this periodic regime, the power balance (\ref{ebilan}) will lead to 
\begin{equation}
 0 = \sum_{\alpha} Q_{\alpha} + W_G + W_{\Gamma}+W_{ch} \ ,
 \label{bilantau}
 \end{equation}
with, in self-explanatory notation ( for the sake of brevity we adopt the notation $<\hat{X}>\  = X$ ), $Q_{\alpha} = \int_{\tau} dt J_{\alpha}^Q(t) $ and $W_i  = \int_{\tau} dt {\mathcal{P}}_{i}(t) $.
Indeed, in the Heisenberg representation, the global density operator is constant leading to $ <\dot{\hat{O}}> = \frac{d <\hat{O}>}{dt}$, thus the left-hand side of Eq. (\ref{ebilan}), when integrated over a period, vanishes.
It is worth noting that not all quantities are periodic in the aforementioned regime. For example integrating $< \hat{\dot{N}}_{\alpha}>$ does not give zero over a period, due to charge transfer 
between the two leads, only the sum over $\alpha$ will cancel. Thus the balance over a period relies also on the fact that there is neither energy or charge accumulation in the dot, nor in the contact regions.

\section{Currents within the Keldysh NEGF}

The Keldysh NEGF technique has been developed in interacting and noninteracting resonant tunneling devices in a famous paper by 
Jauho {\it et al.}~\cite{JauhoWingreenMeir1994}.  This is a powerful technique which is exact in the case of a noninteracting system. 
In absence of Coulomb interaction, another approach, namely the scattering matrix formalism~\cite{Arrachea06}, has been shown to be equivalent to NEGF in the case of periodically driven systems.
In Ref.~\cite{JauhoWingreenMeir1994}, the charge current was derived in the general case.
Since then, expressions for the dynamical energy current have been proposed~\cite{Liu12,crepieux}, however they are restricted to bias or gate voltage modulations. 
We generalize and write down the energy and heat current expressions in the case of time modulation of hybridization parameters too.

We adopt the wide-band limit (WBL) hypothesis, which requires an energy-independent one-spin density of states for the $\alpha$ lead,  $\rho_{\alpha}(\epsilon)$,  as well 
as an energy-independent parameter  $V_{\alpha}(\epsilon)$, such as 
to define a simple hybridization parameter: $ \Gamma^{\alpha} = 2 \pi \rho_{\alpha} |V_{\alpha}|^2$.
In the WBL, by nature, some integrals  are infinite, this has no influence on the balances when integrated along a cycle. 
In the following, we choose the symmetric hybridization case $\Gamma_L = \Gamma_R = {\Gamma}/ 2$.

As first quoted in Eq. (46) of Ref.~\cite{JauhoWingreenMeir1994}, the electric current in the noninteracting resonant-level model reads
\begin{eqnarray}
J_{\alpha }^e (t)  =& -&\frac{2 e }{\hbar}  \Gamma^{\alpha} u_{\alpha}(t) \sum_{\sigma} \int \frac{d \epsilon}{2 \pi}  f_{\alpha}(\epsilon) \mathrm {Im}[  A_{\alpha \sigma}(\epsilon, t)]  \nonumber \\
&- &\frac {e}{\hbar}  \Gamma^{\alpha} \ u_{\alpha} (t)^2\   \sum_{\sigma}n_{\sigma}(t) \ ,
\label{eq:Jelec}
\end{eqnarray}
with, in Jauho's  {\it al.} notations~\cite{JauhoWingreenMeir1994}
\begin{equation}
A_{\alpha \sigma} (\epsilon, t) = \frac 1 \hbar \int dt_1 G^r_{\sigma} (t,t_1) u_{\alpha} (t_1) e^{i \epsilon (t-t_1)/\hbar } e^{-\frac i \hbar \int_t^{t_1} du \  \Delta_{\alpha}(u)} \ .
\end{equation}
 Besides, 
\begin{equation}
G^r_{\sigma} (t,t') = g^r_{\sigma}(t,t') e^{- \frac {1}{ 2 \hbar} \int_{t'}^t dt_1 [ \Gamma^L (u_L(t_1))^2+ \Gamma^R (u_R(t_1))^2  ]} \ ,
\end{equation}
and
\begin{equation}
g^r_{\sigma}(t,t') = - i \theta(t-t') e^{- \frac{i}{\hbar} \int_{t'}^t dt_1 \ \epsilon_d(t_1)} \ .
\end{equation}
$f_{\alpha} (\epsilon) =(e^{(\epsilon-\mu_{\alpha})/T_{\alpha}} +1)^{-1}$ are the lead Fermi functions. Throughout this paper we take $k_B =1$.
Finally, the dot density is evaluated from
\begin{equation}
n_{\sigma}(t)  = \sum_{\alpha} \Gamma^{\alpha} \int \frac{d \epsilon}{2 \pi} f_{\alpha} (\epsilon) |A_{\alpha \sigma} (\epsilon,t)|^2 \ .
\label{density}
\end{equation}

In appendix A we establish the following expression for the energy currents:
\begin{eqnarray}
J^E_{\alpha}(t) = & -&\frac{2  }{\hbar} \Gamma^{\alpha}  \sum_{\sigma} \int \frac{d \epsilon}{2 \pi}  \Bigl[ \  \Bigr] \nonumber \\
&-&\frac{1  }{\hbar}  \Gamma^{\alpha}\ \epsilon_{d}(t ) \  u_{\alpha }(t)^2 \sum_{\sigma}  n_{\sigma}(t)  \nonumber \\
& -&  \dot{\Delta}_{ \alpha } (t) \sum_{k, \sigma}  <c^{\dagger}_{k \alpha \sigma} (t)c_{k \alpha \sigma} (t) > \ ,
\label{eq:Jenerg}
\end{eqnarray}
with 
\begin{eqnarray}
 \Bigl[  \ \Bigr]   =   &\frac 1 2 &  u_{\alpha }(t)^2 \Bigl(   \Gamma^L f_L(\epsilon) u_{L }(t) \mathrm {Re} \bigl[ A_{L \sigma} (\epsilon,t)  \bigr] \nonumber \\
&+&  \Gamma^R f_R(\epsilon) u_{R }(t)  \mathrm {Re} \bigl[ A_{R \sigma} (\epsilon,t) 
   \bigr] 
  \Bigr) \nonumber \\
 &  +& f_{\alpha} (\epsilon) \bigl(\epsilon+\Delta_{\alpha}(t)\bigr) u_{\alpha}(t) \mathrm {Im} \bigl[   A_{\alpha \sigma}(\epsilon,t)    \bigr]  \ .
 \end{eqnarray}
 The heat current is then obtained from Eq. (\ref{heatcurrent}).
 Following Jauho's interpretation~\cite{JauhoWingreenMeir1994} of the charge current (Eq.(\ref{eq:Jelec})) in terms of inflow and outflow 
contributions (respectively first and second terms), we can also detail the energy current (Eq.(\ref{eq:Jenerg})):
the first and second terms are, respectively, the energy counterparts 
of the two charge contributions.  
While the third term in Eq.(\ref{eq:Jenerg}) is the contribution of the work exchanged between 
the lead and the system outside.

 We finally evaluate the mean values corresponding to the different powers supplied by the system outside.
 Two of them, $\mathcal{P}_G(t)$ and $\mathcal{P}_{ch}(t)$ in Eq.(\ref{powers}), are readily evaluated, from the electric currents [Eq.(\ref{eq:Jelec})], and dot density [Eq.(\ref{density})].
 The third one, which is the power exchanged between the whole system and its outside during the hybridization modulation of the $\alpha$-side, can be calculated (see Appendix B) 
 and reads
 \begin{equation}
\mathcal{P}_{\alpha \Gamma} (t)= 2  \Gamma^{\alpha} \ \dot{u}_{\alpha }(t) \sum_{\sigma}  \int \frac{d \epsilon}{2 \pi} f_{\alpha}(\epsilon)\ \mathrm {Re} \Bigl[   A_{\alpha\sigma} (\epsilon,t)  \Bigr] \ .
\label{dotWGam}
\end{equation}
This integral may diverge in the WBL due to the absence of a low-energy cutoff. However, along a cycle, this trouble can be cured by a partial cancellation between the opening and the closing of the hybridization between the dot and the reservoir. 
 Moreover, for ${u}_{\alpha }(t)$ varying as a step function, a $\delta$-like divergence of $\mathcal{P}_{\alpha \Gamma} (t)$ is expected. We shall avoid such abrupt variations in the following.
 
In all calculations, we observed that the sign of this power is opposite to the sign of $\dot{u}_{\alpha}(t)$, such that linking up the dot and the reservoir supplies power to the outside, while closing the connection 
 requires power from the outside. (See later for a numerical example.)
This sign can be understood in a simpler 
model:  suppose a dot is initially disconnected from a lead $\alpha$, the energy for the lead+dot in this initial equilibrium state $E^i$ is readily written. Then the dot and lead are connected. When the new equilibrium state is reached, we can estimate the energy $E^f$ of the whole system: neglecting the dot energy variation, the difference $E^f-E^i $ is equal to $\frac1 2<H_{T \alpha}^{\rm eq}>$ \cite{Bruch15}. This quantity can be evaluated using the single-lead equilibrium expression of $A_{\alpha,\sigma} (\epsilon,t)$,
$ A_{\alpha}^{\rm eq} (\epsilon) = \frac{1}{\epsilon-\epsilon^0+i \Gamma_{\alpha} /2}$, we get 
\begin{equation}
<H^{\rm eq}_{T \alpha}> = -2 \Gamma_{\alpha} \sum_{\sigma} \int \frac{d \epsilon}{2 \pi} f_{\alpha}(\epsilon)\frac{\epsilon^0-\epsilon}{(\epsilon^0-\epsilon)^2+\Gamma_{\alpha}^2 /4} \ ,
\end{equation}
which diverges in absence of any low-energy cutoff, and is negative due to the monotonous decrease of the Fermi function.
 
\section{efficiency and coefficient of performance of a cyclic thermal machine}

The kind of device under scrutiny in this paper offers a great advantage over a mechanical thermal machine:  its great tunability. To switch from an engine to a receptor (refrigerator or heat pump), one needs to invert the mechanical cycle of a usual thermal machine. It is quite simpler for the present thermoelectric devices: one may switch from engine to receptor just by varying the level position with a gate voltage, and/or varying the bias voltage. 
For example, for the static device, with $T_L=30$, $T_R=10$, $\mu_R -\mu_L =60$, all in $\Gamma$ units, changing the dot level from $\epsilon^0=50$ 
to $\epsilon^0=70$ converts a refrigerator to an engine. 
The refrigerator is
characterized by a COP which is about  half that of Carnot, $\mathrm {COP} \simeq  0.54 \ \mathrm {COP}_C$, while the engine is characterized by a relative efficiency of $\eta / \eta_C \simeq 0.8 \  $ ($\eta_C$ is the Carnot efficiency). Such high COP and $\eta$ are close to the best values we can get in such a device for these temperatures,
respectively, $0.56 \ \mathrm {COP}_C$, and $0.83\ \eta_C$.

Let us use the generic term {\it performance} for $\eta$ and COP:
for fixed reservoir temperatures, we have compared various time-driven thermal machines to the static {\it performance}-optimized one~\cite{note2}. We never observed that a better {\it performance} could be found in the time-driven case.  
However, moving away from the optimized static thermal machine, we can enhance {\it performance} by time driving, as will be shown in the next section.

\subsection{Thermoelectric generator (TEG)}

For a cyclic thermal machine (engine or receptor) periodically driven in time, the balance over the period was established in Eq. (\ref{bilantau}).
For a TEG, with $T_L > T_R$, we have $W_{ch} + W_{\Gamma}+W_G <0, Q_L >0$ and $Q_R <0$,
and its efficiency may be defined by
\begin{equation}
\eta = \frac{|W_{ch}+W_{\Gamma}+W_G|}{Q_L}  = 1 - \frac{| Q_R|}{Q_L} \ .
\label{eta}
\end{equation}
Three work contributions come in. 
The chemical work is the useful one, and is related to the charge transferred between the left and right leads. In a forced periodic regime, the
left and right charge currents may be different, however, when the cycle is completed, we have: $ \int_{\tau} dt J^e_L(t) = -\int_{\tau} dt J^e_R(t)$. Indeed, the total charge is conserved and the dot density goes back to its initial value.

The previous definition of efficiency in Eq.(\ref{eta}) may seem not so obvious: indeed, the efficiency may be naturally defined as the ratio between gain and cost. This rather leads to
the expression
\begin{equation}
\eta' = \frac{|W_{ch}|}{Q_L + W_{\Gamma}+W_G} \ .
\end{equation}
However, the latter definition has some drawback compared to Eq.(\ref{eta}): in the context of energy harvesting, we want our machine to be a thermal engine, not a mechanical engine which would use essentially
electric work exchanged through the gate and contact regions to produce electric work. Equation (\ref{eta}) was used also in Ref.~\cite{EspositoPRE}.

\subsection{Refrigerator}

Correspondingly, still with $T_L > T_R$, the coefficient of performance for a refrigerator may be written as
\begin{equation}
{\rm COP} = \frac{Q_R}{W_{ch}+W_{\Gamma}+W_G} = \frac{1}{-\frac{Q_L}{Q_R}-1} \ .
\end{equation}
In the case of a receptor, $W_{ch} +W_{\Gamma}+W_{G} >0$, $Q_L <0$, and $Q_R >0$.

\section{Driven refrigerator}

We illustrate our findings with a setup where the bias voltage is zero. 
Thus, the stationary regime would be totally irrelevant in the context of thermal machines; in presence of a temperature gradient, there would be only a spontaneous heat transfer from the hot reservoir to the cold one.
Instead, in the dynamical regime, we can engineer a refrigerator, by
sequentially connecting and disconnecting the dot from the two reservoirs.
Without loss of generality, by choice of energy origin, we take $\mu_{\alpha}=0$. 

\begin{figure}[h!]
\begin{minipage}[t]{.48\textwidth}
        \begin{center}
            \includegraphics[width=\textwidth]{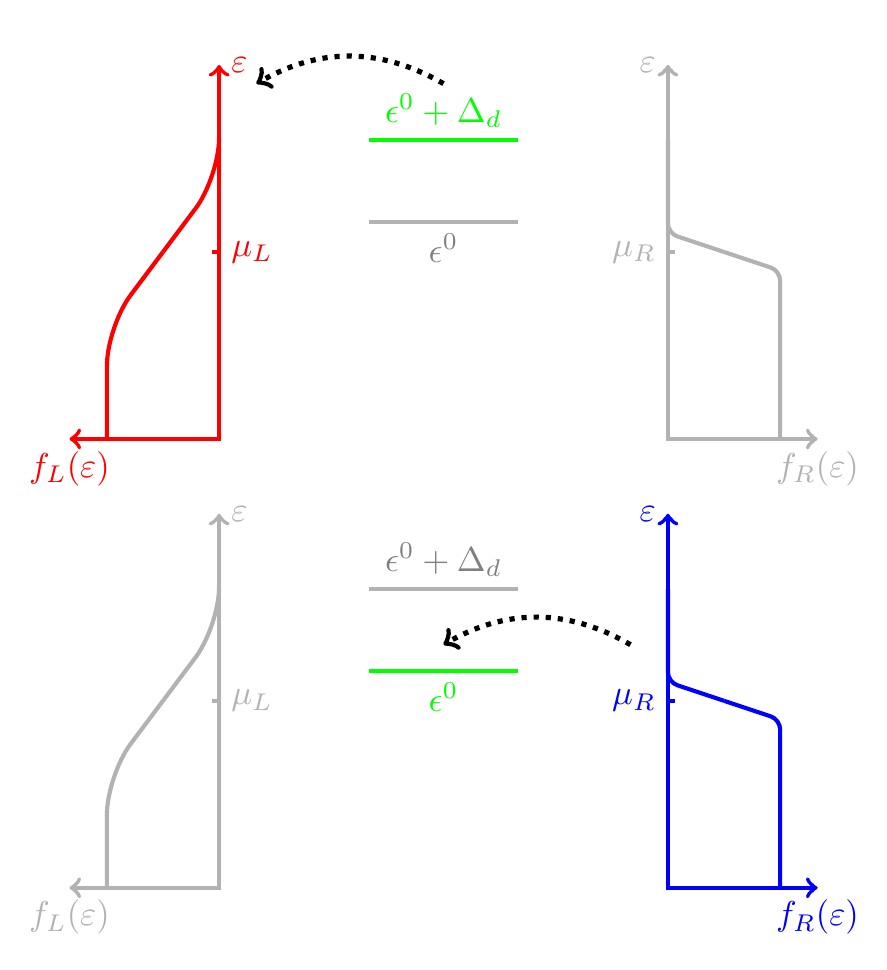}
        \end{center}
    \end{minipage}
    \begin{minipage}[t]{.48\textwidth}
        \begin{center}
            \includegraphics[width=\textwidth]{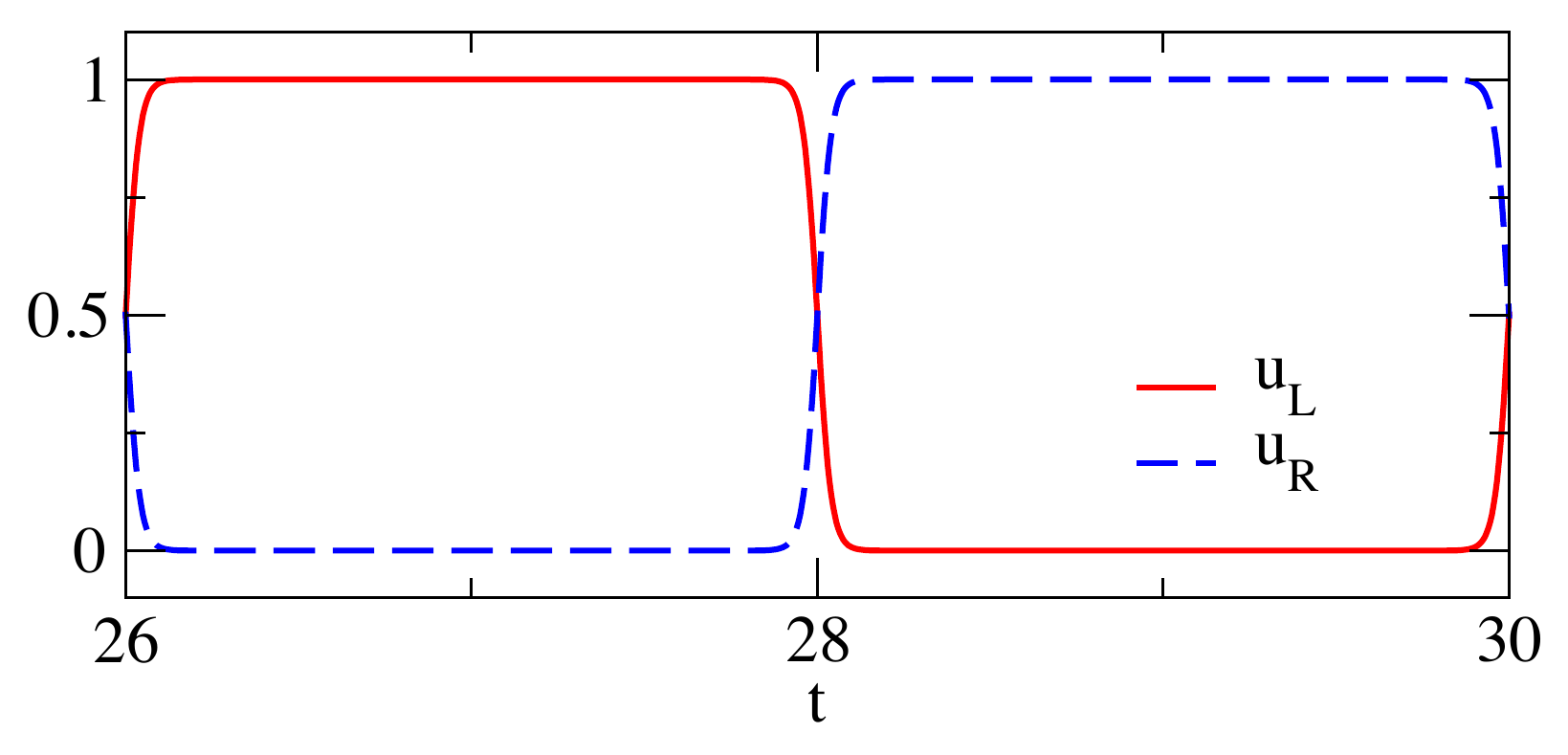}
        \end{center}
    \end{minipage}\\
    \caption{(Color online) Protocol for the driven refrigerator: when the dot level is in the high position, it is connected to the hot reservoir, while when the dot level is low, it is connected to the cold one. Below: time variations of $u_L(t)$ and $u_R(t)$ over a period, $t$ is in $\hbar / \Gamma$ units; for the time range see Ref. \cite{initial}.}
    \label{protoco}
\end{figure}
\begin{figure}[h!]
\begin{center}
\includegraphics[width=.48\textwidth]{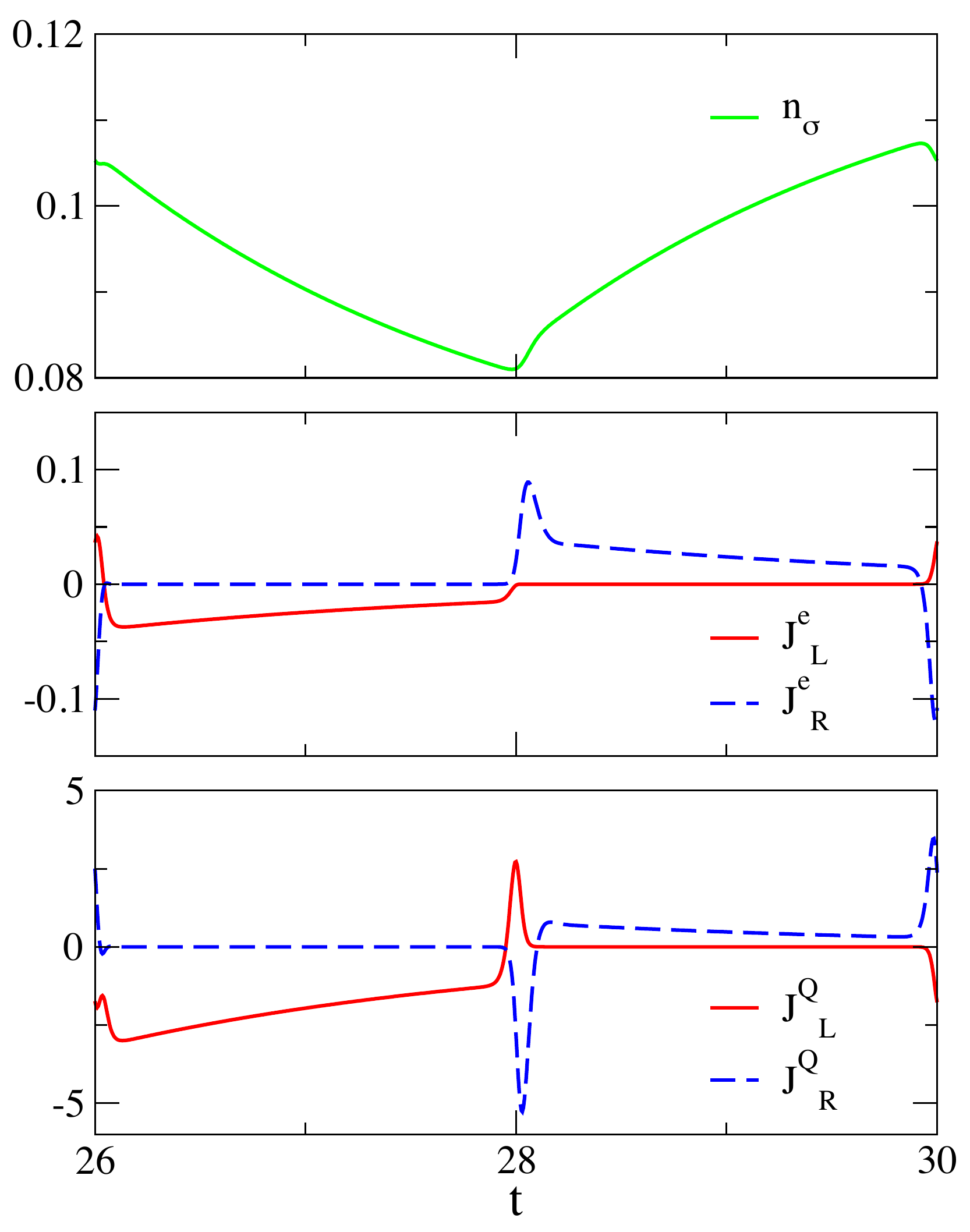}
\caption{(Color online) (Top panel) Dot occupancy per spin. (Middle panel) Electric currents in $ e \Gamma / \hbar$ units. (Bottom) Heat currents in $ \Gamma^2 / \hbar$ units, $t$ is in $\hbar / \Gamma$ units. See text for parameters, and Fig. 2 for driving protocol.}
    \label{Scenario1}
\end{center}
\end{figure}
\begin{figure}[h!]
\begin{center}
\includegraphics[width=.48\textwidth]{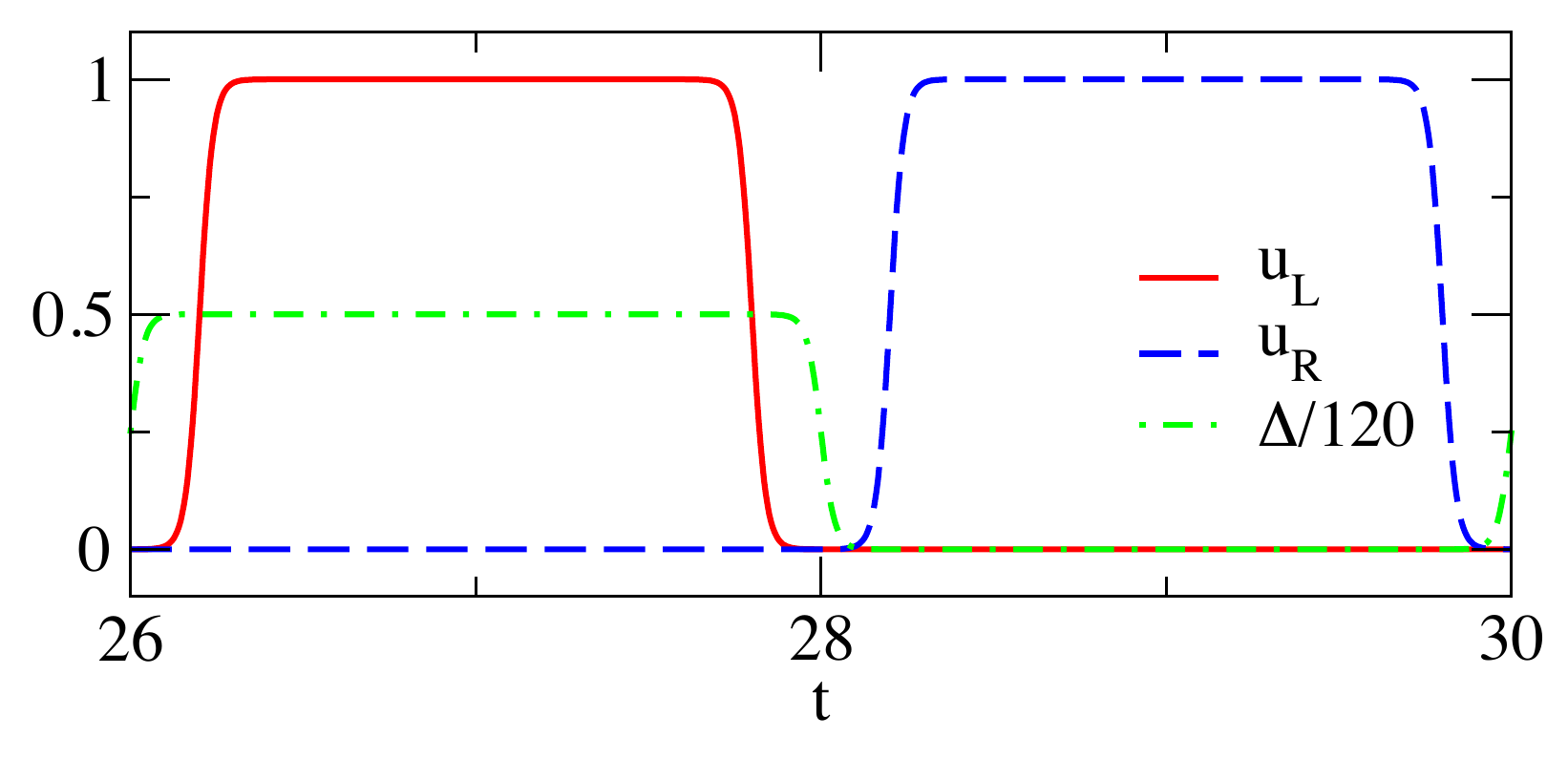}
\caption{(Color online) Protocol for the driven refrigerator in the case of time lags; variations of $u_L(t)$, $u_R(t)$, as well as $\Delta(t)/120$ over a period, [$\Delta(t)$ has been scaled by a factor 120 to match the figure], $t$ is in $\hbar / \Gamma$ units.  Connection and disconnection of leads have been shifted by $0.2 \hbar / \Gamma$ relatively to dot-level modulation.}
    \label{protocol2}
\end{center}
\end{figure}
\begin{figure}[h!]
\begin{center}
\includegraphics[width=.48\textwidth]{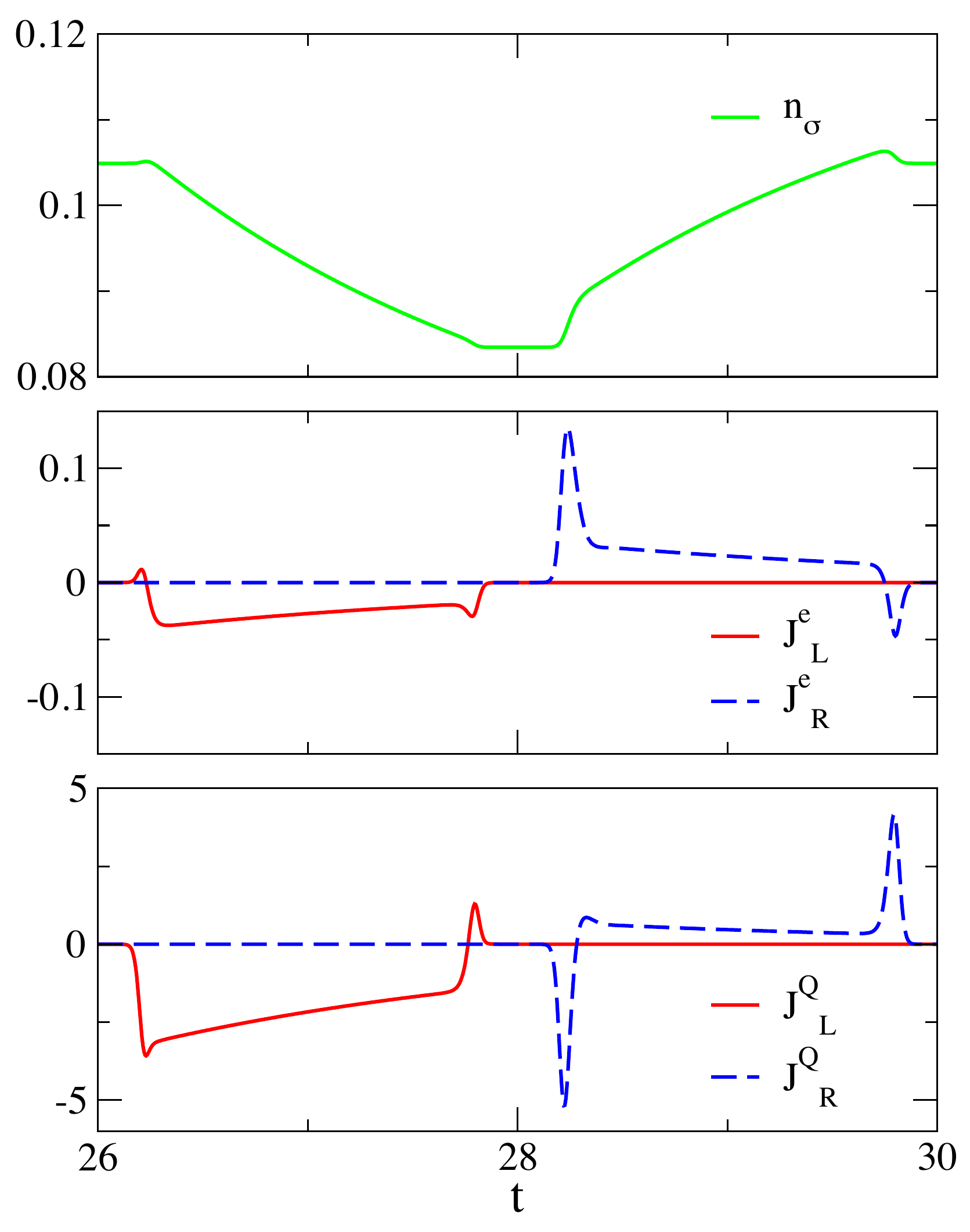}
\caption{(Color online) (Top panel) Dot occupancy per spin. (Middle panel) Electric currents. (Bottom) Heat currents. See text for parameters, Fig. 4 for driving protocol, and Fig. 3 for units.}
    \label{Scenario2}
\end{center}
\end{figure}
\begin{figure}[h!]
\begin{center}
\includegraphics[width=.48\textwidth]{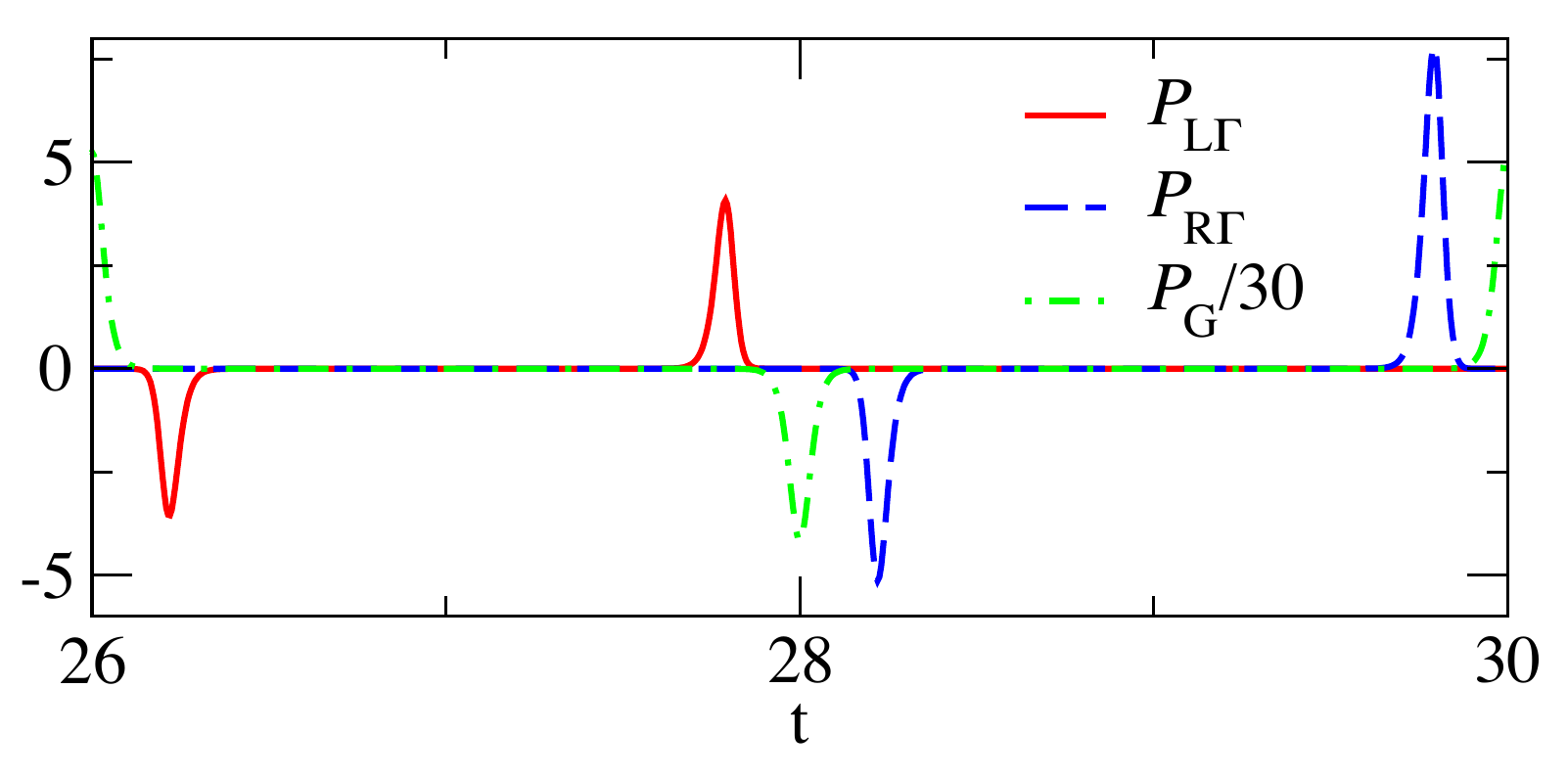}
\caption{(Color online) For the protocol shown in Fig. 4, as defined in text, powers in $ \Gamma^2 / \hbar$ units, $t$ is in $\hbar / \Gamma$ units. $\mathcal{P}_G$ has been reduced by a factor 30 to match the figure.}
    \label{Scenariop}
\end{center}
\end{figure}
As sketched in Fig. 2, in one phase of the cycle, the dot level is in the highest position ($\epsilon^0+\Delta_d$) and is connected to the left hot lead, such that electrons can drop from the
dot to the $L$ lead. Then a second phase begins: the connection to $L$ lead is suppressed, while the connection to the colder $R$ lead opens up, concomitantly with the lowering of the dot level to $\epsilon^0$. In this phase, the dot filling increases. The two phases follow one another periodically. 
A drawing in this figure shows the hybridization modulations $u_L(t)$ and $u_R(t)$, furthermore, we take $\Delta(t)= \Delta_d \times u_L(t)$.
The connections and disconnections from reservoirs, as well as the shift of the dot level, are not exactly step functions to prevent a $\delta(t)$-like divergence of $\mathcal{P}_{\Gamma}(t)$ and of $\mathcal{P}_G(t)$,
but are characterized by quite rapid variations: in a time scale of  about $\tau / 50$, they reach 98\% of the the plateau values, where $\tau = 4 \hbar / \Gamma$ is the cycle period. For definiteness, 
$u_L(t)$ and $u_R(t)$ are composed of step functions 
smoothed by a factor
$(1 - 1/(e^{200(t-t_0)/\tau} +1))$.
In both phases of the cycle, the dot level is higher than the lead chemical potentials. On the grounds of stationary results, a heat current of the same sign as the electric current is thus expected. One can also be convinced about this by taking a look at Fig. 2 and thinking in terms of modification 
of electron distribution in the lead
by electron transfer: a jump to $L$ lead from a dot level higher than $\mu_L$ tends to warm up the left reservoir. On the contrary, 
a jump from the right reservoir onto the dot, from a level higher than $\mu_R$ tends to cool down the corresponding reservoir.
This protocol also leads to charge pumping from the right to the left leads. Note that heat and charge transfers exist also in case of no temperature bias.

Results for the dot density, as well as electric and heat currents are shown in Fig. 3. The parameters are $T_L=30$, $T_R=10$, $\epsilon^0=20$, $\Delta_d=60$, all in $\Gamma$ units. 
One can note that $\Gamma$, the Lorentzian width of the dot level, is  here the smaller energy scale.
The coefficient of performance for the present refrigerator is close to half that of a Carnot refrigerator operating between these reservoirs: we find a COP of 0.24, while the Carnot COP would be 
$1/(T_L/T_R -1)=0.5$.
In more details, for the cycle, one gets~\cite{numericaltest} $Q_L =-3.77\ \Gamma$, $Q_R=0.73\ \Gamma$, $W_{G}=2.90\ \Gamma$, $W_{L \Gamma}= 5 \times 10^{-2}\ \Gamma$, $W_{R \Gamma}= 9 \times 10^{-2} \ \Gamma$, and an average transferred charge per cycle $q=4.6 \times 10^{-2} e$. 
One can notice that the charge is very small due to the combination of two facts. Firstly, during the cycle, the dot level is always above the lead chemical potentials, such that the dot occupancy is low. Secondly, the time driving is done at a rather high frequency.

The charge currents, as well as the heat currents in Fig. 3 display sharp and rapid variations which deserve to be discussed.
To disentangle the effects of connection and disconnection of reservoirs, as well as the shift of the dot level, we consider a slightly modified setup where the modulations are no more concomitant, as shown in Fig. 4.
The delays between opening, closing, and shift only slightly modify the heat and work exchanged over a cycle, indeed, for the former parameters, one gets 
$Q_L =- 3.46\Gamma$, $Q_R= 0.67\Gamma$, $W_{G}= 2.57\Gamma$, $W_{L \Gamma}= 3 \times 10^{-2} \Gamma$, $W_{R \Gamma}= 0.18  \Gamma$, and transferred charge per cycle $q= 4.3 \times 10^{-2} e$. The COP is thus unchanged.

The results for electric currents, density, as well as heat currents are shown  in Fig. 5.
It can be observed that the hybridization shutdown coincides with a dip of the corresponding charge current, whereas the hybridization opening is associated with a fast increase of the electric current entering the dot. However, these currents settle in rapidly, in a too short time to 
be allocated to charge tunneling. Indeed, the rise time or fall time of $u_{\alpha}(t)$ are roughly of the order of one-tenth of the tunneling time 
$\hbar / \Gamma$.  Rather, such inflows and outflows can be attributed to the modification of the dot and lead frontiers: 
after connection the dot and the leads spread, the frontiers are dynamically blurred.
For dot levels well above the chemical potential, setting $\alpha$-hybridization corresponds to a dot inflow. 
Conversely, closing it is accompanied by an outflow.
The situation is reversed for dot levels well below the chemical potential.
Inflows and outflows are corroborated by a perturbative approach, as shown in Appendix C.
The over-currents seen in our results are reminiscent of those described in experiments in Refs.~\cite{Kaestner08,Blumenthal07}, which were ascribed to time-dependent lead-dot couplings and time-dependent dot-level position.

Note that along a cycle there is no compensation between these over-currents, due to a modification of $n_{\sigma}(t)$ between opening and closing. 
After the over-current fading, the expected behavior settles in, however,  
the period is too short to reach the steady state. Besides, due to the $\Gamma$ broadening of the dot level, in the steady state the dot occupancy would differ from the Fermi function value. 

Let us turn to the heat currents, which are shown in Figs. 3 and Fig. 5 (bottom panels), respectively without and with delays between different modulations. The heat over-currents are concomitant with the electric ones, but are of opposite sign: this is another argument against charge tunneling. Their sign is also explained in the perturbative treatment discussed in Appendix C.
Except during these rapid transient behaviors, charge and heat currents have the same direction, as previously discussed for electron transfer beyond chemical potential.
It is worth noting that the over-currents, which are related to fast modulation, would be absent in master equation calculations.

The thermal machine performance depends significantly on the driving protocol. 
However, it is not easy to predict the one which will maximize the COP.
For the refrigerator, we have $|Q_L| > Q_R$,  and a smaller $|Q_L|$ associated 
with a bigger $Q_R$ would improve it.
Looking at Fig. 3, lower panel, it can be seen that over-currents can be beneficial  or 
detrimental to the refrigerator: first, they do not balance between opening 
and closing, and if the net contribution to the heat over the cycle is positive for $Q_L$, 
it is negative for $Q_R$.
This is an advantage for $|Q_L|$, but a disadvantage for $Q_R$. 
Many parameters can be varied: the period, the steepness, the duty cycle for example, but also the dot level positions. We did not make a 
completely comprehensive study, but the protocol detailed in Fig. \ref{protoco} is close to the best  
we found from the COP point of view for the chosen temperatures. It can be slightly improved with a steeper hybridization modulation 
$dl= \tau /400$, associated with a longer period $\tau = 8 \hbar / \Gamma$, this leads to a COP close to $0.28$, which is the best COP of the static machine (see previous section).

In Fig. 6, the power corresponding to a connection or a disconnection between the lead and the dot, as well as the power corresponding to moving the dot level is drawn over a period. 
The present numerical calculations were done with a low-energy cutoff 
equal to $-250 \ \Gamma$ [see the discussion following Eq.(\ref{dotWGam})]. For the
three quantities, the positive peaks follow the negative ones: as if, in some sense, during a cycle, the energy was partly temporarily borrowed and returned between the system and its outside. The $W_{\alpha \Gamma}$ integrals over one period are small but finite,  being an order of magnitude greater than the numerical precision.

Again, in our protocol, where the balance is established between disconnected states, the problem of interpretation of energy extracted from or spilled into the leads disappears. 
For the actual parameters ($\mu_{\alpha}=0$ and $\dot{\Delta}_{\alpha}=0$), from Eq.(\ref{heatcurrent}), $\hat{J}_{\alpha}^Q(t)$ and
$\hat{J}_{\alpha}^E(t)$ coincide. Thus the performance is robust and does not suffer of misinterpretation. 

Except at very low temperature (of the order of, or lower than Kondo temperature), the 
Coulomb interaction would not influence much the thermal machine performance.
Indeed, in the large $U$ case, the major effect is to restrict the channel number from
two to one. Then the charge and heat currents would be two times smaller than
for $U=0$, keeping the COP unchanged. 
In the intermediate case $U \sim \Gamma$, on the grounds of a previous study on electric currents \cite{Vovchenko14}, one can expect some renormalization of the flows, leading to a substantially similar COP. However, this conclusion would deserve further validation. Finally, the noninteracting multi-level 
model, multiplying the channel number, obviously leads to the same performance than the one discussed in the paper.

\section{Conclusion}

We have derived the energy and heat current expressions for a system composed of a dot coupled with two reservoirs. The model parameters, such 
as bias and gate voltages, as well as hybridizations, are modulated in time.

We have applied our results to engineer an efficient refrigerator setup, for which there is no static counterpart. 
To avoid the pitfall of strong coupling, which may lead to an equivocal definition of extracted or injected energy, 
we only consider cycles linking disconnected subsystem states. 

Finally, we have observed transient charge and heat currents which are not of tunneling origin, and we have shown 
that their behaviors, which can be understood in a perturbative approach,
are related to the difficulty of defining boundaries between subsystems. 

\section{Acknowledgment}
We gratefully acknowledge Janet Anders, Michael Moskalets, and Robert S. Whitney for 
helpful comments and suggestions.

\section*{Appendix A: evaluation of the energy current}

Taking  the mean value of Eq.(\ref{currents}), it can be established that 
\begin{eqnarray}
J^E_{\alpha}(t) = && \frac {2}{\hbar} \sum_{k, \sigma} \varepsilon_{k \alpha} (t) \mathrm{Re} \bigl{[} V_{k \alpha}(t) G^<_{d k \alpha \sigma} (t, t)\bigr{]} \nonumber \label{courantene} \\ 
&& - \dot{\Delta}_{ \alpha } (t) \sum_{k, \sigma}  n_{k \alpha \sigma} (t) \ ,
\end{eqnarray}
where the lesser Green's function is defined as
\begin{equation}
G^<_{d k \alpha \sigma} (t, t')= i < c^{\dagger}_{k \alpha \sigma}(t') d_{\sigma}(t)> \ .
\end{equation}
Using the equation of motion and Langreth rules, one can show (\cite{JauhoWingreenMeir1994}, Eq.(12)) that
\begin{eqnarray}
G^<_{d k \alpha \sigma} (t, t')&&= \frac{1}{\hbar} \int  dt_1 G^r_{dd\sigma}(t,t_1) V^*_{k \alpha}(t_1) g^<_{k \alpha}(t_1,t') \nonumber \\
&&+\frac{1}{\hbar} \int  dt_1 G^<_{dd\sigma}(t,t_1) V^*_{k \alpha}(t_1) g^a_{k \alpha}(t_1,t') 
\label{langreth}
\end{eqnarray}
Notations are close to those found in Ref.~\cite{JauhoWingreenMeir1994} and not recalled here. Using this result,
the first line contribution to $J^E_{\alpha}(t)$ in Eq.(\ref{courantene}) can be written as
\begin{equation}
J^E_{\alpha I}(t) = - \frac{2}{\hbar} \sum_{\sigma} \int_{-\infty}^{t} dt_1 \int \frac{d \epsilon}{2 \pi} (\epsilon+\Delta_{\alpha}(t)) \mathrm{Im} \bigl [ \   \bigr ] \ ,
\end{equation}
where
\begin{equation}
\bigl [ \   \bigr ] = \frac{1}{\hbar}\Gamma^{\alpha} (\epsilon, t_1,t) e^{i \epsilon (t-t_1)/\hbar} \Bigl( f_{\alpha}(\epsilon)G^r_{dd\sigma}(t,t_1) + G^<_{dd\sigma}(t,t_1)  \Bigr) \ .
\end{equation}
This equation is quite similar to the electric current expression of Jauho {\it et al.} (\cite{JauhoWingreenMeir1994}, Eq.(15)), except for the energy term $(\epsilon+\Delta_{\alpha} (t))$
in the integrand.  This general expression, valid without the WBL hypothesis and in the case of Coulomb interaction on the dot,  was also given in
Refs.~\cite{crepieux,Ludovico14}, in absence of bias modulation.
We cut $J^E_{\alpha I}(t) = J^E_{\alpha Ir}(t)+J^E_{\alpha I l\Delta}(t)+J^E_{\alpha I l\epsilon}(t)$ in three contributions, the first one arising from $G^r_{dd\sigma}$, the second one from $\Delta_{\alpha}\ G^<_{dd\sigma} $, and
the last one from $\epsilon\ G^<_{dd\sigma} $.
Along the lines followed for the electric current evaluation, the first term in the WBL is readily evaluated, and gives
\begin{eqnarray}
&&J^E_{\alpha Ir}(t) = - \frac{2}{\hbar}  \Gamma^{\alpha} u_{\alpha}(t) \times  \nonumber  \\
&&\int \frac{d \epsilon}{2 \pi} f_{\alpha}(\epsilon) (\epsilon+\Delta_{\alpha}(t))  \sum_{\sigma} \mathrm{Im} \bigl [ A_{\alpha \sigma} (\epsilon, t)\   \bigr ] \ .
\end{eqnarray}
The integral may diverge logarithmically in the WBL (as in the stationary case). 
This divergence will be potentially canceled by another term in the case of dot level modulation, however, it may persist in the case of hybridization connection or disconnection.
The evaluation of the term resulting from $ \Delta_{\alpha}\ G^<_{dd\sigma} $ is also direct, thanks to the integration over $\epsilon$.
We get
\begin{equation}
J^E_{\alpha I l\Delta}(t) = - \frac{1}{\hbar} \Gamma^{\alpha}  \Delta_{\alpha}(t) u_{\alpha}(t)^2\sum_{\sigma}n_{\sigma}(t) \ .
\end{equation}
The third contribution can be evaluated using the derivative of the Dirac function
\begin{equation}
i \int \frac{d \epsilon}{2 \pi} \epsilon \ e^{i \epsilon t /\hbar} = \hbar^2\delta' (t) \   , {\rm with} \ \int dt f(t) \delta' (t) = - f'(0) \ .
\end{equation}
After a lengthy but straightforward calculation, this leads to 
\begin{eqnarray}
J^E_{\alpha I l\epsilon}(t) &=& - \frac{1}{\hbar} \Gamma^{\alpha}  (\epsilon^0+\Delta(t)) u_{\alpha}(t)^2 \sum_{\sigma}n_{\sigma}(t) \nonumber \\
&& + \frac{1}{\hbar} \Gamma^{\alpha}  \Delta_{\alpha}(t) u_{\alpha}(t)^2 \sum_{\sigma}n_{\sigma}(t) \nonumber \\
&-&\frac{1}{\hbar}  \Gamma^{\alpha} u_{\alpha}(t)^2 \sum_{\beta} \Gamma^{\beta} \times \nonumber \\
&& u_{\beta}(t)  \int \frac{d \epsilon}{2 \pi}  f_{\beta} (\epsilon)   \sum_{\sigma}\mathrm{Re} \bigl [ A_{\beta \sigma} (\epsilon, t)\   \bigr ] \ .
\end{eqnarray}
Gathering the different previous contributions, we finally get the announced result quoted in Eq.(\ref{eq:Jenerg}).

\section*{Appendix B: Evaluation of $\mathcal{P}_{\alpha \Gamma} (t) $}

From the definition (Eq.(\ref{powers})), $\mathcal{P}_{\alpha \Gamma} (t) $ can be written in terms of the previously defined Green's function $G^<_{d k \alpha \sigma} (t, t')$:
\begin{equation}
\mathcal{P}_{\alpha \Gamma} (t) = 2 \sum_{k, \sigma} \mathrm{Im} \bigl{[} \dot{V}_{k \alpha}(t) G^<_{d k \alpha \sigma} (t, t)\bigr{]} \ .
\end{equation}
From Eq.(\ref{langreth}), this leads to two contributions. The one related to $ G^<_{d d \sigma}$ leads to a term proportional to $\mathrm{Im} \bigl{[} \frac{\dot{u}_{\alpha}(t)}{u_{\alpha}(t)} \bigr{]}$, which cancels: indeed, as discussed in Ref.~\cite{Splettstoesser05} the phase of the hybridization modulation must be time independent.
The second contribution originating from $G^r_{dd\sigma}$ leads to the expression quoted in Eq.(\ref{dotWGam}).

\section*{Appendix C: Over-current signs}

It can be found in textbooks that in the first-order perturbation theory, the probability of transition from an initial state of 
energy $\epsilon_i$, to a final state of energy $\epsilon_f$,  at short time $t$ ($t \ll \frac{\hbar}{|V_\alpha|}$), after a sudden switch at $t=0$ of a time-independent potential $V_\alpha$, reads
\begin{equation}
\mathcal{P}(i \rightarrow f,t) =4 \frac{\sin^2{\Bigl( (\epsilon_i-\epsilon_f) t / 2 \hbar \Bigr)}}{(\epsilon_i-\epsilon_f)^2} |V_{\alpha}|^2 \ ,
\end{equation}
indicating that at short times, the perturbation can induce transitions between the dot and the lead continuous spectrum up to energy differences of order $ 2 \pi \hbar / t$. 

To incorporate the mean number of electrons in the lead $\alpha$ as well as on the dot, one has to weight the previous probability by the occupation difference $n_\sigma (1-f_\alpha(\epsilon))-f_\alpha(\epsilon)(1-n_\sigma)= n_\sigma-f_\alpha(\epsilon)$. 
Considering a lead with a density of states $\rho_\alpha(\epsilon)$, we thus obtain an evaluation of the charge current in this perturbative approach:
\begin{equation}
J^e_{\alpha \ perturb}(t) =  e \Gamma_\alpha \sum_{\sigma} \int \frac{d \epsilon}{2 \pi} \ 4 \ \frac{\sin^2{(\epsilon_d -\epsilon) t / 2 \hbar}}{t\ (\epsilon_d -\epsilon)^2}  [f_\alpha(\epsilon)-n_\sigma] \ ,
\end{equation}
where we chose the same convention as in the main text for the sign, and 
where we used the WBL relation $\rho_{\alpha} |V_{\alpha}|^2= \frac{\Gamma_\alpha}{2 \pi}$. 
This result predicts the right sign for the electric over-currents at the hybridization opening observed in the main text. Furthermore the agreement is nearly quantitative 
in the case of a steep connection between the dot and the reservoir at short times.
The perturbative approach enables a similar evaluation of the heat current, just by replacing the electric charge in the previous integrand by
$(\epsilon-\mu_\alpha)$. 
It confirms the heat over-current sign observed in the main text, and gives also 
a quantitative agreement for its order of magnitude 
at short times in the case of a steep rising connection.
In the case of disconnection between the dot and the lead, the evaluation of the currents at short times is not so simple due to the spread of the dot level before disconnection. However, one may invoke the time reversal symmetry: all other things being equal, the inflow/outflow at the opening will correspond to the outflow/inflow at the closing.


\begin{thebibliography}{99}

\bibitem{Carnot} S. Carnot, {\it R\'eflexions sur la puissance motrice du feu et sur les machines propres \`a d\'evelopper cette puissance} (Bachelier, Paris, 1824).

\bibitem{Pekola15} J. P. Pekola, Nature Phys. {\bf 11}, 118 (2015). 

\bibitem{Brantut13}  J.-P. Brantut,   C. Grenier,  J. Meineke,  D. Stadler, S. Krinner, C. Kollath, T. Esslinger, and A. Georges, Science {\bf 342}, 713 (2013).

\bibitem{Hicks} L. D. Hicks, and M. S. Dresselhaus, Phys. Rev. B {\bf 47}, 12727 (1993).

\bibitem{Costi10} T. A. Costi and V. Zlati\'c, Phys. Rev. B. {\bf 81}, 235127 (2010).

\bibitem{Azema12} J. Azema, A.-M. Dar\'e, S. Sch\"afer, and P. Lombardo, Phys. Rev. B {\bf 86}, 075303 (2012). 

\bibitem{Donsa14} S. Donsa, S. Andergassen, and K. Held, Phys. Rev. B. {\bf 89}, 125103 (2014).

\bibitem{Nakpathomkun10} N. Nakpathomkun, H. Q. Xu, and H. Linke, Phys. Rev. B {\bf 82}, 235428 (2010).

\bibitem{Whitney14} R. S. Whitney, Phys. Rev. Lett. {\bf 112}, 130601 (2014).

\bibitem{Whitney15} R. S. Whitney, Phys. Rev. B. {\bf 91}, 115425 (2015).

\bibitem{Arrachea07} L. Arrachea, M. Moskalets, and L. Martin-Moreno, Phys. Rev. B {\bf 75}, 245420 (2007).

\bibitem{Rey07} M. Rey, M. Strass, S. Kohler, P. H\"anggi, and F. Sols, Phys. Rev. B {\bf 76}, 085337 (2007).

\bibitem{EspositoEPL2010} M.  Esposito, R. Kawai, K. Lindenberg, and C. Van den Broeck,  Europhys. Lett. {\bf 89}, 20003 (2010).

 \bibitem{Liu12}  W. Liu, K. Sasaoka, T. Yamamoto, T. Tada, and S. Watanabe, Jap. J. Appl. Phys. {\bf 51}, 094303 (2012).
 
\bibitem{crepieux} A. Cr\'epieux, F. \v{S}imkovic, B. Cambon, and F. Michelini  Phys. Rev. B {\bf 83}, 153417 (2011); 
 {\bf 89}, 239907 (2014).

\bibitem{Ludovico14} M. F. Ludovico, J. S. Lim, M. Moskalets, L. Arrachea, and D. S\'anchez, Phys. Rev. B {\bf 89}, 161306(R) (2014).

\bibitem{Ludovico14bis} M. F. Ludovico, J. S. Lim, M. Moskalets, L. Arrachea, and D. S\'anchez, J. Phys: Conf. Ser. {\bf 568}, 052017 (2014).

\bibitem{Zhou15} H. Zhou, J. Thingna, P. H\"anggi, J.-S. Wang, and B. Li, Sci. Rep. {\bf 5}, 14870 92015).

\bibitem{Chirla14}   R. Chirla, and C. P. Moca, Phys. Rev. B {\bf 89}, 045132 (2014).

\bibitem{Brandner15} K. Brandner, K. Saito, and U. Seifert, Phys. Rev. X 5, 031019 (2015).

\bibitem{JauhoWingreenMeir1994}
A.-P.~Jauho, N. S.~Wingreen, and Y.~Meir, Phys. Rev. B {\bf 50}, 5528 (1994).

\bibitem{schmidt08} T. L. Schmidt, P. Werner, L. M\"uhlbacher, and A. Komnik, Phys. Rev. B {\bf 78}, 235110 (2008).

\bibitem{Croy12} A. Croy, U. Saalmann, A. R. Hern\'andez, and C. H. Lewenkopf, Phys. Rev.B {\bf  85}, 035309 (2012).

\bibitem{Vovchenko14} V. Vovchenko, D. Anchishkin, J. Azema, P. Lombardo, R. Hayn, and A.-M. Dar\'e, 
J. Phys.: Condens. Matter {\bf 26}, 015306 (2014).

\bibitem{Eissing15} A. K. Eissing, V. Meden, and D. M. Kennes, arXiv:1508.01325 [cond-mat.str-el].

\bibitem{Kaestner15} For a recent review, see B. Kaestner, and V. Kashcheyevs, Rep. Prog. Phys. 78, 103901 (2015).

\bibitem{EspositoPRE} M. Esposito, R. Kawai, K. Lindenberg, and C. Van den Broeck, Phys. Rev. E {\bf 81}, 041106 (2010).

\bibitem{Juergens13} S. Juergens, F. Haupt, M. Moskalets, and J. Splettstoesser, Phys. Rev. B {\bf 87}, 245423 (2013).

\bibitem{Torfason13}  K. Torfason, A. Manolescu, S. I. Erlingsson, and V. Gudmundsson, Physica E {\bf 53}, 178 (2013).

\bibitem{Ludovico15} M. F. Ludovico, F. Battista, F. von Oppen, and L. Arrachea, arXiv:1506.08617 [cond-mat.mes-hall].

\bibitem{Haupt13} F. Haupt, M. Leijnse, H. L. Calvo, L. Classen, J. Splettstoesser, and M. R. Wegewijs, Phys. Status. Solidi B {\bf 250}, 2315 (2013).

\bibitem{EspositoPRL15} M. Esposito, M. A. Ochoa, and M. Galperin, Phys. Rev. Lett {\bf 114}, 080602 (2015).

\bibitem{Splettstoesser05} J. Splettstoesser, M. Governale, J. K\"{o}nig, and R. Fazio, Phys. Rev. Lett. {\bf 95}, 246803 (2005).

\bibitem{Blumenthal07} M. D. Blumenthal, B. Kaestner, L. Li, S. Giblin, T. J. B. M. Janssen, M. Pepper, D. Anderson, G. Jones, and D. A. Ritchie, Nat. Phys. {\bf 3}, 343 (2007). 

\bibitem{Kaestner08} B. Kaestner, {\it et al.},  Phys. Rev. B {\bf 77}, 153301 (2008).

\bibitem{Gemmer} J. Gemmer, M. Michel, and G. Mahler, {\it Quantum Thermodynamics}, (2nd Ed., Springer, Berlin, 2009).

\bibitem{initial} In the present calculations the initial state is the stationary state corresponding to a level $\epsilon^0$ linked to the two reservoirs, respectively characterized by 
$(T_{\alpha}, \mu_{\alpha})$. After about six periods for the present protocols, the periodical regime is reached. This is the reason for the time range 
shown in the figures.

\bibitem{Arrachea06} L. Arrachea, and M. Moskalets, Phys. Rev. B {\bf 74}, 245322 (2006).

\bibitem{Bruch15} A. Bruch, M. Thomas, S. V. Kusminskiy, F. von Open, ans A. Nitzan arXiv:1511.03276 [cond-mat.mes-hall].

\bibitem{note2} For two fixed temperatures $T_{\alpha}$ in $\Gamma$ units, optimizing from the efficiency or the COP point of view the static thermal machine, is done by exploring the two dimensional space spanned by $\epsilon^0$ and $( \mu_R -\mu_L)$. 

\bibitem{numericaltest} The fulfillment of the balance, see Eq.(\ref{bilantau}), is a test of our numerical calculations. 


\end{thebibliography}
\end{document}